\documentclass[twocolumn,showpacs,floatfix,aps]{revtex4}
\usepackage[dvips]{graphics}

\def \beq {\begin{equation}}
\def \eeq {\end{equation}}

\usepackage{amssymb}

\begin{document}

\title{Overspinning a nearly extreme   black hole and \\
the Weak Cosmic Censorship conjecture}

\author{Maur\'\i cio Richartz}
\email{richartz@ifi.unicamp.br}
\affiliation{ Instituto de F\'\i sica Gleb Wataghin, UNICAMP, \\
C.P. 6165, 13083-970 Campinas, SP, Brazil.}
\author{Alberto Saa}
\email{asaa@ime.unicamp.br}
\affiliation{
Departamento de Matem\'atica Aplicada,
  UNICAMP, \\
C.P. 6065, 13083-859 Campinas, SP, Brazil.}

\begin{abstract}
We revisit here the recent proposal for overspinning a nearly extreme black hole by means
of a quantum tunneling process. We show that electrically
neutral massless fermions
   evade possible
back reactions effects related to superradiance, confirming the view that it would be
indeed possible to
  form a
naked singularity
due to   quantum effects.
\end{abstract}

\maketitle

\section{Introduction}

The Weak Cosmic Censorship conjecture (WCCC) states basically
that any spacetime singularity   originated
after a gravitational collapse   must be hidden inside
an event horizon\cite{WCCC}. The conjecture, which is   believed to be true at
{\em  classical level}, is one of the most important open problems
in General Relativity.
We remind that, without a full description of   spacetime singularities,
  the WCCC must be true in order   to assure
the predictability of the laws of Physics\cite{predic}. Gravitational collapse and
gedanken experiments trying to destroy the event horizons of   black holes  are
common
theoretical tests of WCCC.
According to well known no-hair
theorems\cite{nohair}, all stationary black hole solutions of
Einstein-Maxwell equations are uniquely determined by
three  parameters: the   mass $M$,
the electrical charge $Q$, and the angular momentum $J$,
which satisfy
\beq
\label{wccc}
M^2 \ge Q^2 + (J/M)^2,
\eeq
with the equality corresponding to the case of an extreme black hole (we
adopt here natural units in which $G=\hbar=c=1$).
Solutions for which $M^2 < Q^2 + (J/M)^2$ have no event horizon;
their central singularities are exposed and they are named
  naked singularities\cite{Wald}. Many {\em classical} results\cite{classical}
have established
that it is impossible for a physical process
to increase $Q$ (to overcharge) or $J$ (to overspin)
a black hole in order to violate (\ref{wccc}). Such results have strongly
 supported the belief
  that the WCCC is true at classical level.

Quantum effects, on the other hand,
 have already altered our understanding of the classical
laws of black holes in the past.  Hawking radiation\cite{hawking}, for instance, implies the decreasing
of the black hole area,  a process known to be classically forbidden\cite{Wald}.
This was precisely the main motivation behind the recent work of Matsas and Silva\cite{MS},
where a quantum tunneling process
leading to the overspinning of a black hole
 is proposed. They consider a nearly extreme
Reissner-Nordstrom black
hole ($Q/M \approx 1$ and $J=0$) and show that the probability of absorbing
low energy  massless scalar particles with   high angular momentum is
  non-vanishing. By   conservation  of energy and angular momentum, they conclude
  that (\ref{wccc}) could be violated after the absorption. In particular,
they show  that for a black hole with
$M=100$ Planck mass units and $Q=M-e$, $e = 1/\sqrt{\alpha} \approx 1/\sqrt{137}$ being the elementary
charge in Planck charge units,
a   particle with very low
energy and total angular momentum $L=\sqrt{\ell(\ell+1)}$, with $\ell = 413$,
 would be enough to
overspin  the black hole and produce a naked singularity.
Larger nearly extreme Reissner-Nordstrom black holes would require  larger total angular momentums
in order to produce   naked singularities. In fact, $\ell\sim M^{3/2}$ for large $M$.

As Matsas and Silva stress in their paper,
the transfer of a high amount  of angular momentum to an initially static black
hole   raises some doubts about the role of possible  back-reaction effects in such   kind of
process. This is   the  central
point of Hod's   contribution\cite{Hod} to   Matsas and Silva's process: when
a wave (or particle) with large angular momentum  approaches a black hole,
higher order back reaction interactions could trigger the rotation of the black hole before the
tunneling. Hod shows then  that superradiance effects would imply that only
those modes with frequency $\omega$ and azimuthal number $m$ such that
\beq
\omega > m\Omega,
\eeq
where $\Omega$ is the angular velocity of the black hole, could be really
absorbed by the black hole, leading eventually to the conclusion
 that it is impossible for such a
process to increase $J$ without increasing $M$ simultaneously, preserving
(\ref{wccc}) and saving the WCCC. We remind that superradiance was proposed by Misner as
a version for waves of the Penrose process  to extract energy from a rotating
black-hole.
Superradiance  is known   to affect in   similar ways
 (bosonic) fields of spins $s=0$ (scalar), $s=1$ (photons)
and $s=2$ (gravitons)\cite{Chandra}. Besides,
there is   an equivalent of the   superradiance effect  for charged situations\cite{bekens},
implying  that it is also impossible to overcharge a black hole by the absorption  of
 low energy scalar charged particles.

In this paper, we go a step further in this problem
by showing that it is possible to reduce arbitrarily the  back reaction
effects raised by Hod\cite{Hod}  by considering the quantum tunneling of neutral  fermions   into
nearly extreme rotating black holes, implying
that quantum effects could indeed lead to the appearance of naked singularities.
Physically, the validity of the process presented here rests on the well known fact that there
is no superradiance effect for electrically neutral massless fermions\cite{fermions}.
We notice also that some recent observations suggest that
rapidly spinning black holes could be rather common in the universe\cite{rapid}.

\section{Fermions around black holes}

The Dirac equation is known to be separable in a Kerr-Newman spacetime\cite{PageD}.
By using Boyer-Lindquist coordinates, a neutral massless Dirac fermion   can
be separated in terms
 of the modes
\beq
\label{modes}
u_{s\omega\ell m}(t,r,\theta,\phi) =  e^{-i\omega t}  R_{s\omega\ell m}(r)S_{s\omega\ell m}(\theta) e^{im\phi},
\eeq
where the radial $R_{s\omega\ell m}(r) =  R$ and angular $S_{s\omega\ell m}(\theta) = S$
functions satisfy
the (spin weight  $s=\pm\frac12$) Teukolsky equations\cite{PageD}
 \begin{widetext}
\begin{eqnarray}
\label{req}
\Delta^{-s} \frac{d}{dr} \left(\Delta^{s+1}  \frac{dR }{dr}  \right) +
\left(
\frac{K^2 -  2is (r-M)K}{\Delta} + 4is \omega r + 2am\omega -a^2\omega^2 - \lambda
\right)R   &=&0, \\
\frac{1}{\sin\theta} \frac{d}{d\theta} \left( \sin\theta \frac{dS }{d\theta}\right) +
\left[ \left(
a\omega\cos\theta - s\right)^2 - \left( \frac{ s\cos\theta + m }{ \sin \theta}\right)^2
- s(s-1) + \lambda
\right]S  &=& 0,
\label{teq}
\end{eqnarray}
\end{widetext}
with $\Delta = r^2 - 2Mr +a^2 +Q^2$,   $a=J/M$, and $K =(r^2+a^2)\omega - am$, where
$\omega$, $\ell\ge \frac12$, and $-\ell \le m \le \ell$
are, respectively,
the   frequency of the
mode, and the spheroidal   and azimuthal spin-weighted harmonic indexes.
In the limit $a\omega \ll 1$, the angular dependence of (\ref{modes}) reduces to
  the spin weighted spherical harmonic  $_s Y_\ell^m(\theta,\phi)=S_{s\omega\ell m}(\theta)e^{i m\phi}$,
  with corresponding eigenvalues $\lambda = (\ell-s)(\ell+s+1)$.

The low energy sector $M\omega \ll 1$ of the modes (\ref{modes}) can be considered analogously to the
scalar case\cite{Chandra,Star}.
The field configuration associated to the tunneling   of   fermions
into the
black hole
corresponds to the physical boundary conditions of purely ingoing modes at the event
horizon $r_+ = M + \sqrt{M^2-a^2-Q^2}$,
\beq
\label{inghor}
R_{r\rightarrow r_+} \approx \Delta^{-s}e^{ -i(\omega-m\Omega) r^*},
\eeq
and a mixture of both ingoing and outgoing modes at infinity
\beq
R_{r\rightarrow \infty} \approx Y_{\rm in}^{(s)} e^{-i\omega r^*} r^{-1} + Y_{\rm out}^{(s)}
  e^{ i\omega r^*} r^{-(2s+1)}
\eeq
where $r^*$ is the usual tortoise coordinate, defined as $dr^* = (r^2+a^2) dr/\Delta$.
In contrast to the scalar ($s=0$) case,
the calculation of the
 transmission ${\cal T}_{s\omega\ell m}$ and reflection ${\cal R}_{s\omega\ell m}$ coefficients,
 which obey $| {\cal T}_{s\omega\ell m}|^2 + | {\cal R}_{s\omega\ell m}|^2 = 1$,
 is rather tricky since it involves $s$-dependent normalization factors which are different for ingoing and outgoing
 modes.  In particular, we do not have  simply
     $|{\cal T}_{s\omega\ell m}|^2 =1 - |Y_{\rm out}^{(s)}/Y_{\rm in}^{(s)}|^2 = 1/|Y_{\rm in}^{(s)}|^2$
as we do for the $s=0$ case. However, a property of the solutions of (\ref{req}) and
 (\ref{teq})
 discovered  by Teukolsky and Press\cite{TeukoPress} leads to
 \beq
 \label{TTT}
|{\cal T}_{s\omega\ell m}|^2 =  1 - \left|
 \frac{Y_{\rm out}^{(s)}Y_{\rm out}^{(-s)}} {Y_{\rm in}^{(s)}Y_{\rm in}^{(-s)}}\right|,
 \eeq
  facilitating our task considerably.

Incidentally,
the transmission coefficient (\ref{TTT}) has been already calculated by Page\cite{Page1}
in the limit of  low frequency  modes ($M\omega \ll 1$)
  for
the case $Q=0$ in the context of   particle emission by black holes.
  We have
\begin{eqnarray}
\label{T}
\left|{\cal T}_{s\omega\ell m} \right|^2 &=&
\left[
\frac{(\ell-s)!(\ell+s)!}{(2\ell)!(2\ell+1)!!}
 \right]^2 \times \\ 
 &&  \prod_{n=1}^{\ell+1/2} \left[1+\left(\frac{\omega -m\Omega}{n\kappa - \frac{1}{2}\kappa} \right)^2
  \right]\left(\frac{A\kappa}{2\pi} \omega \right)^{2\ell + 1}, \nonumber 
\end{eqnarray}
for $s=\pm\frac12$,
where $A$ stands for the black hole area and $\kappa$ to its surface gravity.
The quantity (\ref{T}) corresponds to the
probability that a mode (\ref{modes}) with a (low) frequency $\omega$, spheroidal
and azimuthal   numbers $\ell$ and $m$, respectively,  be absorbed by the black hole. As one can see,
it is positive for arbitrary small values of $\omega$, in contrast to the bosonic cases where
  superradiance effects are present\cite{superradiance}.

\section{Overspinning the black hole}

Now, we can finally show how to overspin a black hole by means of
the quantum tunneling of fermions, evading
Hod's back reaction issues\cite{Hod}.
Let us suppose, first,
  we have the ``nearest extreme'' possible Kerr black hole:  mass $M$ (in Planck units),
angular momentum $J=M^2-1$, and electrical charge $Q=0$.
This black hole can absorb a mode with arbitrarily small frequency $\omega$
and $\ell = m = 3/2$.
The non-vanishing
 probability for this process
 is
\beq
\label{TT}
\left|{\cal T}_{\frac12 \omega\frac{3}{2}\frac{3}{2}} \right|^2 =
\frac{( M  \omega )^4}{36}\left(1 + 8\frac{a^2}{M^2} \right),
\eeq
valid
in the limit    $M\omega \ll 1$.  From the conservation laws,   after the tunneling  the
black hole will have mass $M_f=M+\omega$, angular momentum $J_f=J+m=M^2+1/2$, and
electrical charge $Q_f=0$. It would be
enough to choose
\beq
\label{omega}
\omega < M \left( \sqrt{1 + \frac{1}{2M^2}} - 1 \right) \sim \frac{1}{4M}
\eeq
 in order to violate (\ref{wccc}) and induce the formation of a naked singularity.
 The last term  in (\ref{omega}) corresponds to the dominant term for large $M$.
In contrast with
Matsas and Silva's original process\cite{MS}, the total amount of angular
momentum that must be transferred to the black hole in order to form a naked singularity
does not depend on $M$ and, moreover, is small.
 A rotating black hole with large mass $M$ and large angular momentum $J = M^2-1$ is not
 expected to be significatively disturbed by a infalling wave with very small frequency $\omega$
 and $\ell = m = 3/2$.
 By choosing large values of $M$,
and consequently large values of $J$, we can
reduce arbitrarily any issue related to rotation higher order interactions raised by
Hod in \cite{Hod}. By dealing with large black holes we can also avoid any complication due to
possible interactions between the infalling wave  and the emitted  Hawking radiation\cite{hawking}.

Nevertheless, we can improve  even more our result by   considering
 a slightly charged nearly extreme    rotating black hole.
Let us add, for instance, a   charge  of $Q=1$   Planck
unit
(corresponding to about
$12 \,e)$
to the black hole with large mass $M$ and angular momentum $J=M^2-1$. Such a black hole,
which will not be
 discharged by Schwinger pair production processes\cite{Schwinger},
 can
be converted into a naked singularity by absorbing a single low energy fermion with minimal
angular momentum!
Since electrically neutral fermions do not couple directly to the black hole
electric  field,
the transmission coefficient
for low frequencies modes   in a   Kerr-Newman black hole
is given essentially by the same Page's formula (\ref{T}),
leading to the following non-vanishing probability for the tunneling of   a  neutral massless   fermion
with low frequency $\omega$ and $\ell = m = 1/2$
\beq
\label{T1}
\left|{\cal T}_{ \frac12\omega\frac{1}{2}\frac{1}{2}} \right|^2 =
(M^2-Q^2)\omega^{2},
\eeq
valid
in the limit    $M\omega \ll 1$.
Due to the
conservation laws, after the absorption the black hole will
have mass $M_f=M+\omega$, angular momentum $J_f=M^2-1/2$, and charge $Q_f=1$.
In order to violate (\ref{wccc}) and produce a naked singularity,
$\omega$ must be chosen as
\beq
\omega < M \left(\sqrt{ \frac{1+\sqrt{1+4\left(M^2-\frac{1}{2}\right)^2}}{2M^2}} - 1 \right) \sim
\frac{1}{16M^3},
\eeq
where
the last term also corresponds to the dominant term for large $M$.
This process could lead to  the
formation of a naked singularity by the tunneling  of a
single neutral fermion with {\em minimal} angular momentum    into a slightly charged
nearly extreme rotating black hole with arbitrary mass $M$. Again, by
considering large black hole masses one can minimize all the
back reactions effects raised by Hod.

There is, nevertheless, a remaining  possible source of back reaction effects in our scenario:
the sudden disappearance
of the event horizon due to the particle tunneling. A complete description of these effects
would certainly require  a full quantum gravity theory. In spite of that,
 some string theoretical results do indeed suggest
that the disappearance of the event horizon could be a
smooth process without, consequently, any sudden back reaction effect
that could alter significantly our conclusions.
In the $D-$brane picture used by Maldacena to describe near
extremal black holes\cite{Maldacena} the (essentially non-perturbative)
 vanishing of the event horizon   is
mapped into a well defined perturbative process of a
low energy effective string model. In particular, the
absorption and emission rates of photons and fermions by nearly extreme
black holes do not show any evidence of sudden phenomena\cite{coeff}.

Although the probabilities of the absorption for both cases considered here are
extremely small, we have shown that it would be indeed possible {\em in principle} to   overspin
a nearly extreme black hole through the quantum
tunneling of low energy fermions. Moreover, we have shown that the process proposed
here, in contrast  to Matsas and Silva's original one\cite{MS}, involves the transfer of
a minimal amount of
angular momentum   from the   quantum field to the black hole.
Since we deal with neutral fermion fields, which are known to be free from superradiance
effects, and manipulate only minimal amounts of energy and angular momentum, we can evade
all  the back reaction issues pointed out by Hod\cite{Hod}, rendering
the overspinning of a black hole by quantum effects a quite robust conclusion.

Spacetime singularities belong
 naturally to the realm of quantum gravity.
We believe that only a complete quantum gravity theory will be able to describe
 naked singularities properly,
dissecting them conclusively
or even restoring the WCCC in a more fundamental level.

\acknowledgements
The authors are grateful to G. Matsas, R.A. Mosna, and C.A.S. Maia
for enlightening discussions.
This work was supported by FAPESP and CNPq.


\begin{references}
\bibitem{WCCC}R. Penrose, Riv. Nuov Cimento {\bf 1}, 252 (1969);
 in {\em General Relativity, an Einstein
Centenary Survey}, Ed. S.W. Hawking and W. Israel, Cambridge University
Press (1969);
C. J. S. Clarke, Class. Quantum Grav. {\bf 11}, 1375 (1994);
R.M. Wald, {\em Gravitational Collapse and Cosmic Censorship},
gr-qc/9710068.
\bibitem{predic} S.W. Hawking, Phys. Rev. D {\bf 14}, 2460 (1976).
\bibitem{nohair} M. Heusler, {\em Black Hole Uniqueness Theorems}, Cambridge University
Press (1996).
\bibitem{Wald} R.M. Wald, {\em General Relativity}, The University of
Chicago Press  (1984).
\bibitem{classical}  C. V. Vishveshwara, Phys. Rev. D {\bf 1}, 2870 (1970);
R. Price, Phys. Rev. D {\bf 5}, 2419 (1972);   Phys.
Rev. D {\bf 5}, 2439 (1972);
 B. S. Kay and R. M. Wald, Class. Quant. Grav. {\bf 4}, 893
(1987);
 B. F. Whiting, J. Math. Phys. {\bf  30}, 1301 (1989);
 D. G. Boulware, Phys. Rev. D {\bf  8}, 2363 (1973);
  R. M. Wald, Ann. Phys. {\bf 82} , 548 (1974);
  T. Needham, Phys. Rev. D {\bf 22}, 791 (1980); S. Hod, Phys. Rev. D {\bf 60}, 104031 (1999);
  Phys. Rev. D {\bf 66}, 024016 (2002).
  \bibitem{hawking} S.W. Hawking, Nature  {\bf  248}, 30 (1974); Comm.
Math. Phys. {\bf 43}, 199 (1975).

\bibitem{MS} G.E.A. Matsas and A.R.R. da Silva, Phys. Rev. Lett. {\bf 99}, 181301 (2007).
\bibitem{Hod} S. Hod, Phys. Rev. Lett. {\bf 100}, 121101 (2008).

\bibitem{Chandra}   S. Chandrasekhar, {\em The Mathematical Theory of Black Holes},
Oxford University Press (1998).



\bibitem{bekens} J.D. Bekenstein, Phys. Rev. D {\bf 7}, 949 (1973).


\bibitem{fermions} W. Unruh, Phys. Rev. Lett. {\bf 31}, 1265   (1973);
R. Guven, Phys. Rev. D {\bf 16}, 1706    (1977);
 B. R. Iyer and Arvind Kumar, Phys. Rev. D {\bf 18}, 4799 (1978);
 S. M. Wagh and N. Dadhich, Phys. Rev. D {\bf 32}, 1863   (1985).

 \bibitem{rapid}J.-M. Wang,  Y.-M. Chen, L.C. Ho, and R.J. McLure, Astrophys. J. {\bf 642}  L111, (2006);
 R. S. Nemmen, R. G. Bower, A. Babul, T. Storchi-Bergmann,
 Month. Not.   R. Astr. Soc. {\bf 377}, 1652 (2007).

\bibitem{PageD} S.A. Teukolsky, Phys. Rev. Lett. {\bf 29}, 1114 (1972); Astrophys. J.
{\bf 185}, 635 (1973);
D. Page, Phys. Rev. D {\bf 14}, 1509   (1976).

\bibitem{Star}A.A. Starobinsky and S.M. Churilov, Sov. Phys. JETP {\bf 38}, 1 (1974).

\bibitem{TeukoPress} S.A. Teukolsky and W.H. Press, Astrophys. J. {\bf 193}, 443 (1974).

\bibitem{Page1} D. Page, Phys. Rev. D {\bf 13}, 198   (1976).

\bibitem{superradiance} Compare the following formulas of \cite{Page1}: (13),  valid for bosons ($2s$ even,
 Hod's\cite{Hod} formula (9) corresponds to $s=0$), and (14) (our (\ref{T})),
valid for fermions.

\bibitem{Schwinger} W.T. Zaumen, Nature {\bf 247}, 530 (1974);
G.W. Gibbons, Comm. Math. Phys. {\bf 44},  245 (1975).

\bibitem{Maldacena} J. Maldacena, Phys. Rev. D {\bf 55}, 7645 (1997).

\bibitem{coeff} J. Maldacena and A. Strominger, Phys. Rev. D {\bf 56} 4975 (1997); S.S. Gubser
Phys. Rev. D {\bf 56} 7854 (1997).


\end{references}
\end{document}